# NEW DEVELOPMENTS IN LINEAR COLLIDERS FINAL FOCUS SYSTEMS

P. Raimondi, A. Seryi, SLAC, Stanford, CA

*Abstract*

The length, complexity and cost of the present Final Focus designs for linear colliders grows very quickly with the beam energy. In this paper, a novel final focus system is presented and compared with the one proposed for NLC [1]. This new design is simpler, shorter and cheaper, with comparable bandwidth, tolerances and tunability. Moreover, the length scales slower than linearly with energy allowing for a more flexible design which is applicable over a much larger energy range.

## 1 INTRODUCTION

The main task of a linear collider final focus system (FFS) is to focus the beams to the small sizes required at the interaction point (IP). To achieve this, the FFS forms a large and almost parallel beam at the entrance of the final doublet (FD) which contains two or more strong quadrupole lenses. For the nominal energy, the beam size at the IP is then determined by $\sigma = \sqrt{\varepsilon \beta^*}$ where $\varepsilon$ is the beam emittance and $\beta^*$ is the betatron function at the IP (typically about 0.1-1mm). However, for a beam with an energy spread $\sigma_E$ (typically 0.1-1%), the beam size is diluted by the chromaticity of these strong lenses. The chromaticity scales as $L^*/\beta^*$, where $L^*$ (typically 2-4 m) is the distance from the IP to the FD, and thus the chromatic dilution of the beam size $\sigma_E L^*/\beta^*$ is very large. The design of a FFS is therefore driven primarily by the necessity of compensating the chromaticity of the FD.

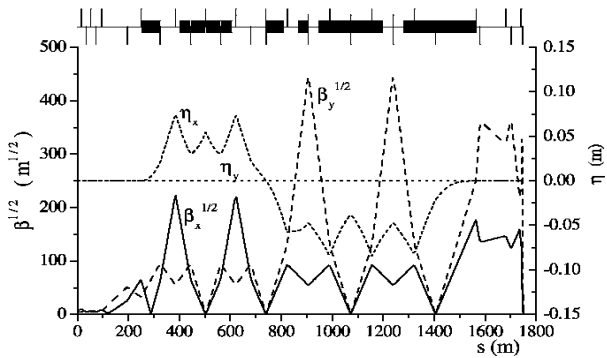

Figure 1: Optics of the traditional Final Focus for the NLC showing horizontal and vertical betatron and dispersion functions. Focusing and defocusing quadrupoles are indicated as up and down bars on the magnet plot above the optics, bends are centered.

In an "traditional" final focus system (SLC [2], FFTB [3] or the new linear collider designs) the chromaticity is compensated in dedicated chromatic correction sections (CCX and CCY) by sextupoles placed in high dispersion and high beta regions. The geometric aberrations generated by them are canceled by using them in pairs with an identity transformation between them. As an example, the "traditional" design of the NLC Final Focus [1] with $L^* = 2m$, $\beta_x^* = 10mm$ and $\beta_y^* = 0.12mm$ is shown in Fig.1. The advantage of the traditional FFS is its separated optics with strictly defined functions and straightforward cancellation of geometrical aberrations. This makes such a system relatively simple for design and analysis.

The major disadvantage of the "traditional" final focus system is that the chromaticity of the FD is not locally compensated. As a direct consequence there are intrinsic limitations on the bandwidth of the system due to the unavoidable breakdown of the proper phase relations between the sextupoles and the FD for different energies. This precludes the perfect cancellation of the chromatic aberrations. Moreover, the system is very sensitive to any disturbance of the beam energy in between the sources of chromaticity, whether due to longitudinal wake-fields or synchrotron radiation. In particular, the bends in the system have to be long and weak to minimize the additional energy spread generated. In addition, the phase slippage of the off-momentum particles drastically limits the dynamic aperture of the system. Therefore very long and problematic collimation sections are required in order to eliminate these particles that would otherwise hit the FD and/or generate background in the detector.

The collimation section optics itself also becomes a source of aberrations since very large beta and dispersion functions are required.

As a result of all these limitations, the length of the beam delivery system becomes a significant fraction of the length of the entire accelerator, and scaling to higher energies is difficult.

## 2 "IDEAL" FINAL FOCUS SYSTEM

Taking into account the disadvantages of the traditional approach, one can formulate the requirements for a more "ideal" final focus:

1) The chromaticity should be corrected as locally as possible.
2) The number of bends should be minimized.
3) The dynamic aperture or, equivalently, the preservation of the linear optics should be as large as possible.
4) The system should be as simple as possible.
5) The system should be optimized for flat beams.

It is straightforward, starting from the IP, to build such a system:

1) A Final Doublet is required to provide focusing.

2) The FD generates chromaticity, so two sextupoles interleaved with these quadrupoles and a bend upstream to generate dispersion across the FD will locally cancel the chromaticity.
3) The sexupoles generate geometric aberrations, so two more sextupoles in phase with them and upstream of the bend are required.
4) In general four more quadrupoles are needed upstream to match the incoming beta function (see the schematic in Fig.2).

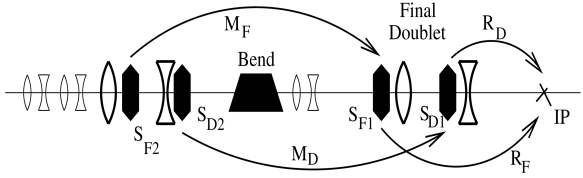

Figure 2: Optical layout of the new final focus.

The second order geometric aberrations are cancelled when the x and y-pairs of sextupoles are separated by transfer matrices $M_F$ and $M_D$:

$$M_F = \begin{vmatrix} F & 0 & 0 & 0 \\ F_{21} & 1/F & 0 & 0 \\ 0 & 0 & F & 0 \\ 0 & 0 & F_{43} & 1/F \end{vmatrix}; \quad M_D = \begin{vmatrix} D & 0 & 0 & 0 \\ D_{21} & 1/D & 0 & 0 \\ 0 & 0 & D & 0 \\ 0 & 0 & D_{43} & 1/D \end{vmatrix}$$

Where all nonzero parameters are arbitrary. In order to cancel the second order chromatic aberrations, the sextupole integrated strengths $K_S$ have to satisfy the equations:

$$K_{SF2} = -F^3 K_{SF1} \qquad K_{SD2} = -D^3 K_{SD1}$$
$$K_{SF1} = \frac{\xi_{x1} + \xi_{x2}}{R_{F12}^3 \eta}, \qquad K_{SD1} = \frac{\xi_y}{R_{D34}^3 \eta}, \qquad (1)$$
$$\xi_{x1} = \xi_{x2} \qquad \xi_x = \frac{d^2 x}{dx' \, dE/E}$$

$x$ and $x'$ are the beam coordinates at the IP, $\xi_{x1}$ is the horizontal chromaticity of the system upstream of the bend, $\xi_{x2}$ is the chromaticity downstream, $\xi_y$ is the vertical chromaticity. $R_F$ and $R_D$ are the transfer matrices defined in Fig.2. The angular dispersion at the IP, $\eta'$, is necessarily nonzero in the new design, but can be small enough that it does not significantly increase the beam divergence. Half of the total horizontal chromaticity of the whole final focus must be generated upstream of the bend in order for the sextupoles to simultaneously cancel the chromaticity and the second order dispersion.

The horizontal and vertical sextupoles are interleaved, so in general they can generate third order geometric aberrations according to:

$$U_{1222} = K_{SD} K_{SF} R_{D12}^2 R_{F12}^2 \varphi_{12}$$
$$U_{3444} = K_{SD} K_{SF} R_{D34}^2 R_{F34}^2 \varphi_{12}$$
$$U_{1244} = -\frac{1}{2} K_{SD} K_{SF} \cdot$$
$$\left[ \left( R_{D34}^2 R_{F12}^2 + R_{D12}^2 R_{F34}^2 \right) \varphi_{12} - 4 R_{D12} R_{D34} R_{F12} R_{F34} \varphi_{34} \right]$$
$$U_{3224} = U_{1244}$$

$\varphi_{12}$ and $\varphi_{34}$ are the elements of the transfer matrix between $S_{F1}$ and $S_{D1}$.

The beam spot sizes dilutions from $U_{3444}$ and $U_{1222}$ are small if the last quadrupole is defocusing and given the typical flat beam parameters like in Tab.1. $U_{1244}$ and $U_{3224}$ can be made to vanish by properly choosing the transfer matrices between the sextupoles. Similar constraints hold for third order chromo-geometric aberrations. All these constraints can be satisfied with the simple system described above. A system with the same demagnification as the NLC FF and comparable optical performance can be built in a length of about 300m.

Table 1: Beam parameters

| | | |
|---|---|---|
| Beam energy, GeV | | 500 |
| Normalized emittances | $\gamma\varepsilon_x / \gamma\varepsilon_y$ ($\mu$m) | 4 / 0.06 |
| Beta functions | $\beta_x / \beta_y$ at IP (mm) | 9.5 / 0.12 |
| Beam sizes | $\sigma_x / \sigma_y$ at IP (nm) | 197 / 2.7 |
| Beam divergence | $\theta_x / \theta_y$ at IP ($\mu$rad) | 21/23 |
| Energy spread | $\sigma_E$ (10$^{-3}$) | 3 |
| Dispersion | $\eta'_x$ at IP (mrad) | 5.4 |

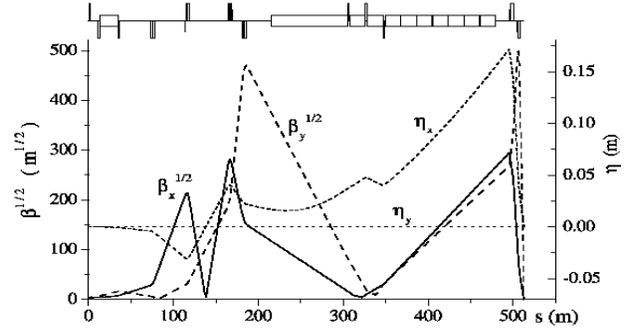

Figure 3: Optics of the new NLC Final Focus System showing horizontal and vertical betatron and dispersion functions similarly to Fig.1

## 3 BANDWIDTH

The new FF system has potentially much better performance than the traditional design. The "minimal" optics concept can be further improved by adding more elements to minimize residual aberrations. An additional bend upstream of the second sextupole pair decreases chromaticity through the system. An additional sextupole upstream and in phase with the last one further reduces third order aberrations in x-plane. Such system has vanishing aberrations up to third order, the residual higher order aberrations can be further minimized by using decapoles. In particular the fourth order aberrations generated by the interleaved horizontal and vertical sextupoles can be reduced with a decapole placed near the closest quadrupole to the IP. The new optics is shown in Fig.3. The flat beam parameters are given in Tab.1. The new system has an $L^*$=4.3m, which is more than twice the original value. This allows the use of large bore superconducting quadrupoles and simplifies the design of the detector. Although the chromaticity is doubled due to the larger $L^*$, the performance of the system is still better than for the original NLC FF design.

Figs. 4a/b compare the bandwidth of the NLC FF and the new design in the IP phase. Figs.5 show the bandwidth in the FD phase. The bandwidth is derived from the variation of the beta function and the beam sizes as they actually contribute to luminosity, which is determined by tracking. The beam size bandwidth is narrower than the beta function bandwidth because of higher order cross-plane chromatic aberrations. While the IP bandwidth for these two systems is comparable, the FD bandwidth is much wider for the new FF.

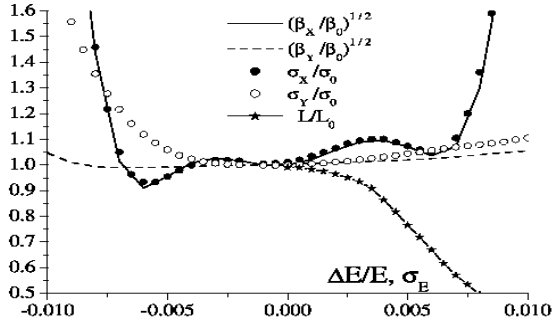

Figure 4a: IP bandwidth of the traditional NLC Final Focus. Normalized beta functions and normalized luminosity equivalent beam sizes Vs energy offset $\Delta E/E$, and normalized luminosity Vs rms energy spread $\sigma_E$.

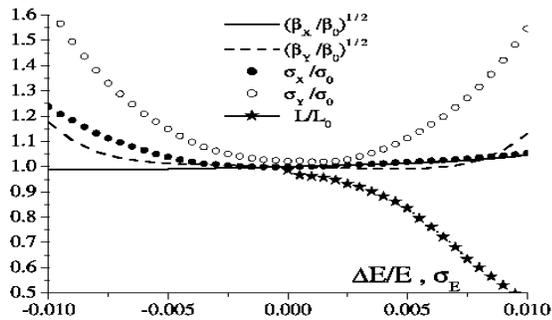

Figure 4b: IP bandwidth of the New NLC Final Focus with definitions as in Fig. 4

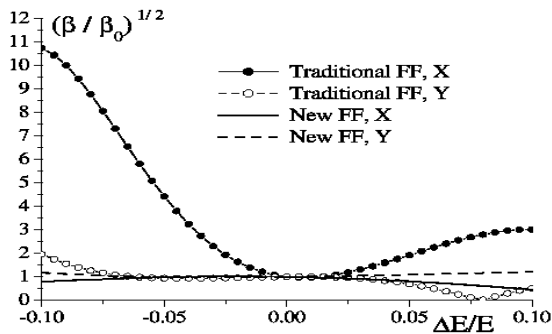

Figure 5: FD bandwidth of the Traditional and New NLC Final Focus. Normalized betatron functions at the final doublet versus energy offset $\Delta E/E$.

Figs.6a/b show the chromaticity through the two systems. The new FF one is much smaller and goes through much fewer optical elements and much shorter distance. This greatly benefits the chromatic properties of the new system and more than compensate for the disadvantage of having the horizontal sextupole pair interleaved with the vertical pair.

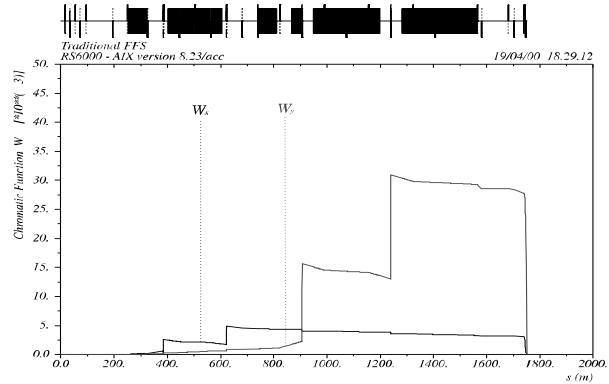

Figure 6a: Horizontal and vertical chromaticity through the NLC Final Focus

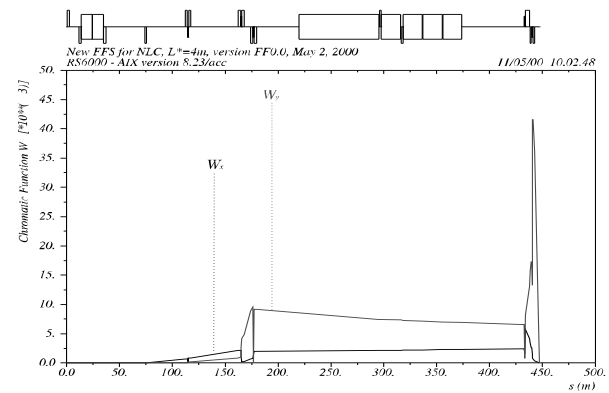

Figure 6b: Horizontal and vertical chromaticity through the New Final Focus

## 4 BACKGROUND

The chromatic aberrations as one of the main sources of the background in the detector have been extensively investigated in SLC. As an example, Fig.7 shows the background in the SLD drift chamber as a function of the chromatic aberration in the doublet face $T_{226} = \dfrac{d^2 x'}{dx' dE/E}$

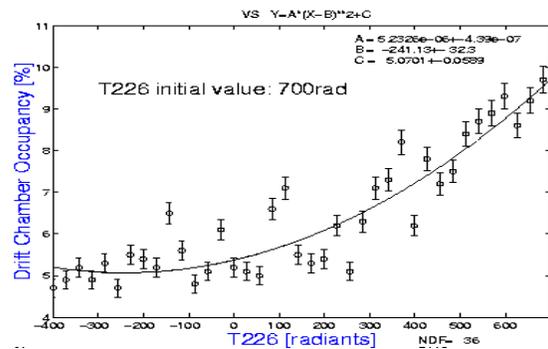

Figure 7: Background in the SLD drift chamber Vs the chromatic aberration $T_{226}$

The design value for this aberration was 700rad, later additional sextupoles were added to minimize it, reducing the background of about a factor 2. Many more high order chromatic aberrations were generated by the sextupoles in the chromatic correction section. It has been observed both in the SLC-FFS and in FFTB that simply turning the sextupoles off, eliminates most of the background and the beam loss monitors signals in the whole beam line.

For SLD-SLC the background was one of the major limiting factors in achieving high luminosity. For future colliders this problem is greatly enhanced: design parameters require beam currents nearly thirty times larger than in SLC at ten times the energy.

A rough estimate of the background could be made using the following assumption:

1) The background is mainly determined by chromo-geometric high order aberrations originated in the CCS's as observed in SLC, for a given aberration it should scale roughly as:

$$Background \propto N\sigma_x^n(FD)\sigma_y^m(FD)\sigma_x^p(IP)\sigma_y^q(IP)(dE/E)^r$$

Being $n$ $m$ $p$ $q$ and $r$ determined by the order and phase of the aberration, $N$ is the beam charge

Assuming similar aberrations as in the SLC-FF, the smaller beam sizes across the FD than what SLC had and the increased beam current, we should expect about 3 times more background than in SLC for a given relative beam collimation.

2) The radiation levels generated in the collimation section should be comparable to the SLC ones. This requirement comes simply by the requirement that the area should maintain residual radiation levels acceptable for human intervention. The possible relative collimation then cannot greatly exceed just about a $3 \cdot 10^{-3}$ of what done at SLC.

From 1) and 2) the expected background could then be about a thousand times higher than in SLC. Clearly this requires a formidable improvement in the collimation section altogether or a minimization of the sources of background in the FFS.

The new FF offers a possible solution to this problem. Fig.8 shows the halo particle distribution at the face of the final doublet for the traditional FF and for the new FF. The beam is very distorted in the traditional FF, very similarly the SLC-FFS, while the nonlinear terms are still negligible for the new FF.

The nonzero dispersion across the FD in the new system has little affect on the dynamic aperture. In addition, the design aperture of the NLC final doublet is about $r_a$=10mm while for the new FF with twice longer $L^*$ this aperture can be as large as $r_a$=40mm. Therefore the collimation requirements for the new FF may be relaxed by a factor of at least one hundred in the IP phase, and by a factor of at least 3 for the FD phase and energy without increasing particle losses at the FD.

Due to the shorter length of the system, there would also be less regeneration of the beam halo in the final focus itself from beam-gas scattering, reducing an additional source of background.

Given the fewer elements and less bends, it could also be possible to build the system with a larger bore aperture everywhere (40mm), further improving the beam stay-clear and the vacuum in the system.

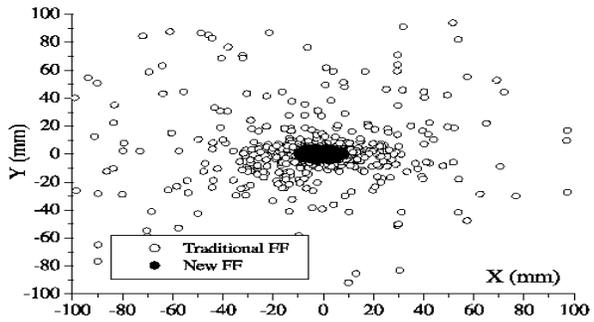

Figure 8: Beam at the entrance of the final doublet for the traditional NLC FF and for the new FF. Particles of the incoming beam are placed on a surface of an ellipsoid with dimensions $N_\sigma(x,x',y,y',E) = (800,8,4000,40,20)$ times larger than the nominal beam sizes.

## 5 TOLERANCES AND TUNING

The effects of magnet displacements on IP beam offsets for the NLC-FF and the new FF are shown in Figs.9a/b respectively. The two systems are relatively similar, most of the contribution to the beam offset at the IP is caused by the FD motion. Fig.10 show the effect of a particular model of ground motion on the vertical beam offset at the IP for the two systems. The main contribution comes from the FD and probably it could be much smaller in the new FF because the longer $L^*$, thus allowing a more rigid support for the magnets.

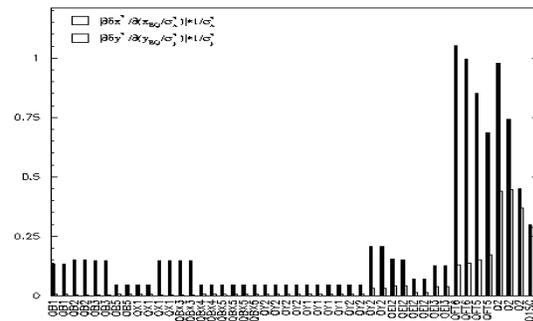

Figure 9a: Traditional FF horizontal (black bars) and vertical (white bars) beam offset at the IP for every magnet unitary displacement (computed with FFADA).

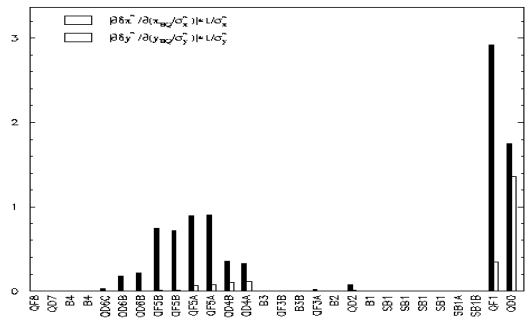

Figure 9b: New FF horizontal (black bars) and vertical (white bars) beam offset at the IP for every magnet unitary displacement (computed with FFADA).

The tuning of the new FFS is very much similar to the old schemes. The main first order aberrations come from quadrupole and skew quadrupoles components generated by sextupole offsets and can be routinely minimized by optimizing spot sizes and luminosity by "knobs" built with sextupole movers. In general for the new FF the luminosity dilutions in the horizontal plane are about 50% worst, since the larger chromaticity required (eq. 1) in the system, however the dominant dilutions come from the vertical plane and those are better mainly due to the simpler scheme with less elements. Higher order spurious aberrations, due to unavoidable lattice errors, can be minimized with additional magnets placed in convenient locations.

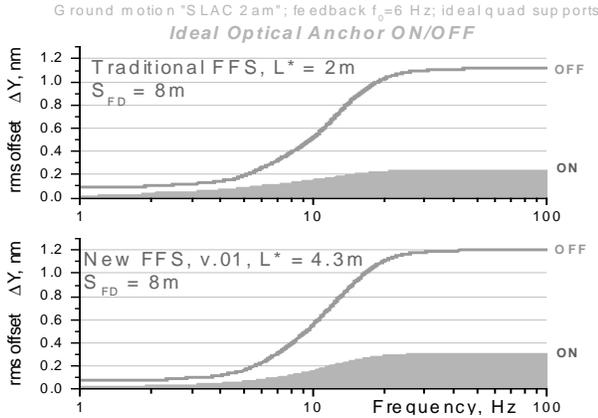

Figure 10: Integrated spectral contribution to the rms vertical beam offset at the IP from ground motion for the NLC-FFS and the new-FFS. The main contribution being from the final doublet (Ideal Optical Anchor OFF)

# 6 SCALING WITH EMITTANCE AND ENERGY.

To maintain optimal performance of the system with larger incoming beam emittances, the bend field must increase like $B_0 \propto \sqrt{\varepsilon}$. The increased field is necessary to hold constant the contribution of high order aberrations to the IP beam size, as well as the contribution of the IP angular dispersion $\eta'_{IP}\sigma_E$ to the beam divergence.

The dependence of the luminosity on beam energy is shown in Fig.11. A fixed length FFS has a wide range of energies where it could operate, especially if the bend field is rescaled.

Scaling to higher energies is very favorable with the new design. For a wide range of parameters, the IP spot size dilution is dominated by the energy spread created by synchrotron radiation in the bends. This scales like:

$$\frac{\Delta\sigma_y}{\sigma_y} \propto \frac{\gamma^5}{L^2}\eta'^3_B \propto (\gamma\varepsilon_y)^{3/2}\left(\frac{\eta'_B L}{\eta'_{IP}}\right)^3\left(\frac{\eta'^2_{IP}}{\varepsilon_y}\right)^{3/2}\frac{\gamma^{7/2}}{L^5}$$

$\eta'_B$ is the angular dispersion produced by the bends, the bend length is assumed to be proportional to the total length of the system $L$. The terms in the parenthesis are constant if the IP angular dispersion is proportional to the beam divergence and if we conservatively assume that the normalized emittance will be the same at higher energies. In this case the length of the system scales with energy as $L \propto \gamma^{7/10}$. If, however, the achievable normalized emittance scales approximately inversely with energy, as is assumed in [4], then the scaling is $L \propto \gamma^{2/5}$. In this case, with the new design, the FF for a 3 TeV center of mass energy collider could be only about 500 m long.

The beam also emits synchrotron radiation in the quadrupoles, which becomes more of a problem at higher energies. This can be reduced in the new design because the larger bandwidth allows the FD quadrupoles to be lengthened to minimize the synchrotron radiation they generate. For the presented optics, the dependence of the luminosity on beam energy is shown in Fig.11. If the beam parameters from [4] are assumed, this FF can operate almost up to 5TeV center of mass energy.

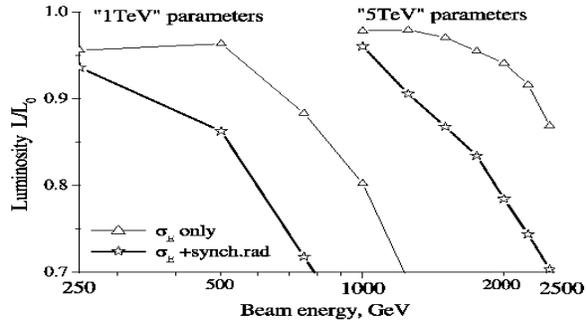

Figure 11: Luminosity Vs beam energy for the new FF, bend field optimized at each energy, with and without synchrotron radiation. The "1TeV" parameters correspond to Tab.1, the "5TeV" set corresponds to [4] with: $\gamma\varepsilon_{x/y} = 50/1\cdot10^{-8}$m, $\beta^*_{x/y}=9.5/0.14$mm, $\sigma_E=0.2\%$, $\sigma_{x/y}$(2.5TeV/Beam)=31/0.54nm.

# 7 CONCLUSION

We have developed a new Final Focus system that has better properties than the systems so far considered and built. It is much shorter, providing a significant cost reduction for the collider. The system has similar bandwidth and several orders of magnitudes larger dynamic aperture. This reduces the backgrounds and relaxes the design of the collimation section. It is also compatible with an $L^*$ which is twice as long as that in the traditional NLC FF design, which simplifies engineering of the Interaction Point area. Finally, its favorable scaling with beam energy makes it attractive for multi-TeV colliders. We believe that further improvements of the performance of the system are possible.

# 8 BIBLIOGRAPHY

[1] NLC ZDR Design Group, "A Zeroth-Order Design Report for the Next Linear Collider", SLAC Report-474 (1996).

[2] J.J. Murray et al., "The Completed Design of the SLC Final Focus System", IEEE Proceedings, (1988).

[3] J. Irwin et al., "The optics of the Final Focus Test Beam",IEEE Proceedings, New York, (1991), p. 2058.

[4] J.P. Delahaye, et al., "A 30 GHz 5-TeV Linear Collider"', PAC Proceedings, (1997), p. 482.